\documentclass[12pt]{article}
\usepackage[dvips]{graphicx}
\usepackage{pdproc}

\usepackage{amsmath,amssymb}

\newcommand{\vev}[1]{\langle {#1} \rangle}

\newcommand{\del}{\partial}

\newcommand{\LS}{\ \ \ \ \ \ \ \ \ \ }
\newcommand{\ls}{\ \ \ \ \ }
\newcommand{\wt}{\widetilde}
\newcommand{\wh}{\widehat}

\newcommand{\ol}{\overline}

\newcommand{\dps}{\displaystyle}
\newcommand{\kahler}{K\"{a}hler }

\newcommand{\bsubeq}{\begin{subequations}}
\newcommand{\esubeq}{\end{subequations}}

\renewcommand{\d}{{\rm d}}

\newcommand{\CP}[2]{{\bf W}{\mathbb C}{\bf P}^{#1}_{#2}}

\renewcommand{\P}[1]{{\mathbb C}{\bf P}^{#1}}

\newcommand{\e}{{\rm e}}

\newcommand{\s}{\footnotesize}

  \makeatletter 
  \def\@cite#1{[#1]} 
  \makeatother    
\begin{document}

\renewcommand{\thefootnote}{\alph{footnote}}

\title{
Towards Mirror Symmetry on Noncompact Calabi-Yau Manifolds
}

\author{ TETSUJI KIMURA}

\address{ 
Theory Division, Institute of Particle and Nuclear Studies,\\
High Energy Accelerator Research Organization (KEK)\\
Tsukuba, Ibaraki 305-0801, Japan
\\ {\rm E-mail: tetsuji@post.kek.jp}}

\abstract{
We study one class of linear sigma models
and their T-dualized theories for noncompact 
Calabi-Yau manifolds.
In the low energy limit,
we find that this system has 
various massless effective theories with orbifolding symmetries.
This phenomenon is new and 
there are no analogous structures in the models for compact Calabi-Yau manifolds
and for line bundles on the simple toric varieties.
}

\normalsize\baselineskip=15pt

\section{Introduction and Summary}

\allowdisplaybreaks{

We constructed metrics on noncompact Calabi-Yau (CY) manifolds 
in the framework of ${\cal N}=2$ supersymmetric nonlinear sigma models \cite{HKN}.
These manifolds are canonical line bundles on compact 
Einstein-\kahler manifolds,
whose mirror pairs have not been well-known yet.
In order to understand the mirror pairs, 
we first study canonical line bundles on
hypersurfaces of projective spaces, the proto-types of
\cite{HKN}, and their mirror pairs in terms of the linear sigma model
\cite{W93} and its T-dualized theory \cite{HV00}.
First we discuss the linear sigma model.
If the Fayet-Iliopoulos (FI) parameter is positive, 
the linear sigma model reduces to a supersymmetric nonlinear sigma model  
whose target space is a canonical line bundle on the degree $k$
hypersurface of $\P{N-1}$, i.e., a noncompact CY manifold.
When we set the FI parameter to be negative, 
there appear various kinds of orbifolded theories as massless effective theories.
Next we study the low energy limit of the T-dualized theory of the linear sigma model.
We obtain two kinds of Landau-Ginzburg (LG) theories with superpotentials including 
positive and negative powers of twisted chiral superfields.
Furthermore we construct two kinds of orbifolded noncompact CY geometries,
which are related to the above two LG theories. 

\section{Linear Sigma Model and its T-dualized Theory}

Let us first consider the linear sigma model \cite{W93} whose 
field configuration and chiral superpotential 
are as follows:\footnote{We omit the explicit expression of the
  Lagrangian,  see Ref. \cite{W93}.}
\begin{gather}
\text{\s $\dps
\begin{array}{c|ccccc} \hline
\text{chiral superfield} & S_1 & \cdots & S_N & P_1 & P_2 \\ \hline
\text{$U(1)$ charge} & 1 & \cdots & 1 & -k & - N +k
\\ \hline
\end{array}
\ls
\begin{array}{l}
\wt{W} \ = \ - \Sigma \, t \; , \\
W_{\rm LSM} \ = \ P_1 \cdot G_k (S_i) \; , 
\end{array}
$}
\label{config}
\end{gather}
where
$S_i$, $P_1$ and $P_2$ are charged chiral superfields and $\Sigma$ is
a twisted chiral superfield coming from the gauge multiplet; 
$t$ is a complex parameter combined with the FI parameter $r$ and the
Theta angle $\theta$ such as $t= r - i \theta$; 
$G_k (S_i)$ is a quasi-homogeneous polynomial of degree $k$ written by
$N$ chiral superfields $S_i$.\footnote{For simplicity we set 
$2 \leq k \leq N-1$.} 
The quasi-homogeneity insists that 
when $G_k (S)$ and its derivatives $\del_i G_k (S)$ vanish 
the variables $S_i$ are all zero.
The above field configuration (\ref{config}) satisfies 
the ``CY condition'' such as $\sum_a Q_a = N - k - (N-k) = 0$.
Due to this, the FI parameter does not receive
one-loop renormalization effects 
and we find that this system is free from 
the UV divergence through perturbative calculations.
The potential energy density for the scalar component fields is given by
\begin{align}
\begin{split}
\text{\s $U$} \ &\text{\s $=$} \ 
\text{\s 
$\dps \frac{e^2}{2}  
\Big\{ r - \sum_{i=1}^N |s_i|^2 + k |p_1|^2 + (N-k) |p_2|^2 \Big\}^2
+ \big| G_k (s_i) \big|^2 
+ |p_1|^2 \cdot \sum_{i=1}^N \big| \del_i G_k (s_j) \big|^2
$}
\\
\ & 
\text{\s $\dps 
\ \ \ + 2 |\sigma|^2 \Big( \sum_{i=1}^N |s_i|^2 
+ k^2 |p_1|^2 + (N-k)^2 |p_2|^2 \Big)
\; .
$}
\end{split}
\label{pot-OCPNk}
\end{align}

Let us analyze classical 
supersymmetric vacuum manifolds $U =0$ and massless effective theories on them.
When $r>0$,
we find that the vacuum manifold ${\cal M}_{\rm CY}$ is 
nothing but the canonical line bundle on the degree $k$
hypersurface of projective space represented as ${\cal O}(-N+k) \to \P{N-1}[k]$.
In the IR limit $e\to \infty$, 
the linear sigma model reduces to the ${\cal N}=2$ supersymmetric
nonlinear sigma model on ${\cal M}_{\rm CY}$.

In $r <0$, the shape of the vacuum manifold drastically changes 
to a very complicated one.
Furthermore, 
there exist four massless effective theories on this vacuum space
expressed in Figure \ref{vacuum-mfd},
\begin{figure}[h]
\begin{center}
\includegraphics[height=5.5cm]{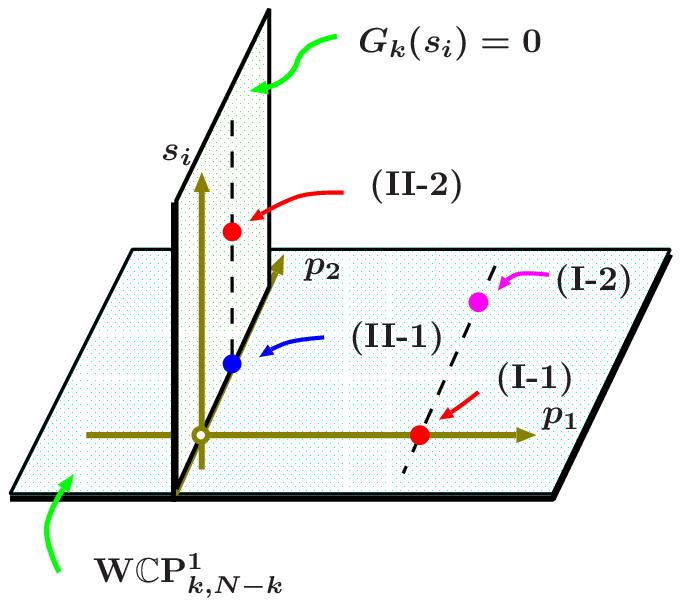}
\ \ \ 
\includegraphics[height=6.1cm]{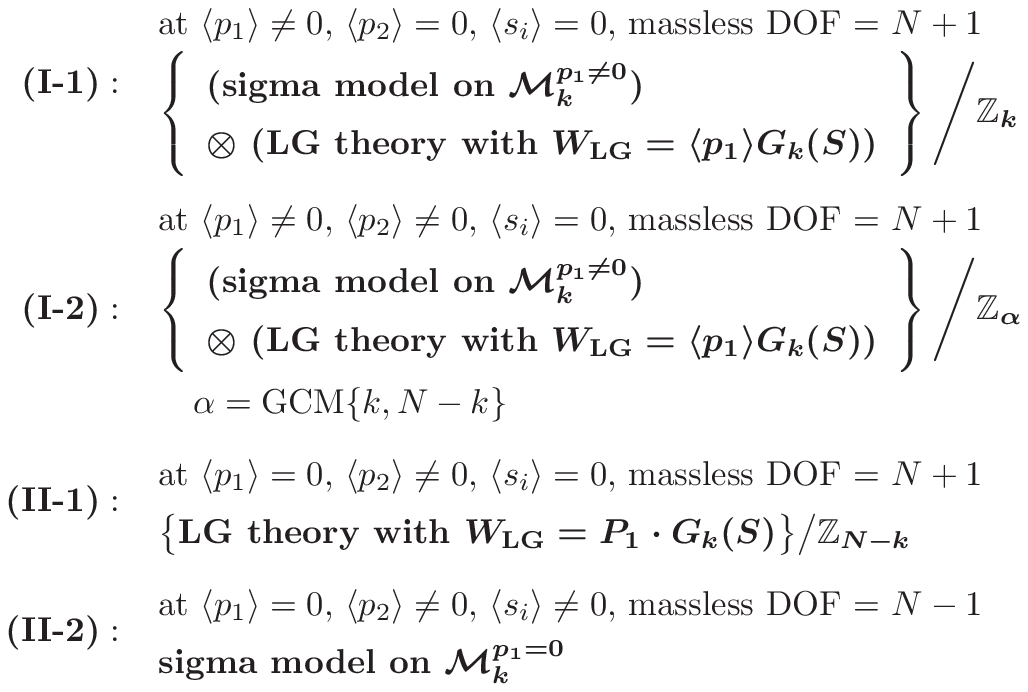}
\end{center}
\caption{Four massless effective theories on the vacuum manifold, when $r < 0$.}
\label{vacuum-mfd}
\end{figure}
where $\CP{1}{k,N-k}$, ${\cal M}_k^{p_1 \neq 0}$ and ${\cal M}_k^{p_1 =0}$ 
are defined in (\ref{mfds}):
\begin{align}
\begin{split}
\text{\s $\CP{1}{k,N-k}$} \ &\text{\s $=$} \ 
\text{\s $\dps 
\Big\{ (p_1, p_2) \in {\mathbb C}^2 \, \Big| \,
r = - k |p_1|^2 - (N-k) |p_2|^2 
\Big\} 
\; , \ \ \ 
{\cal M}_k^{p_1 \neq 0} = \big\{ \CP{1}{k,N-k} \, \big| \, p_1 \neq 0 \big\}
\; , $} \\
\text{\s ${\cal M}_k^{p_1 =0}$} \ &\text{\s $=$} \
\text{\s $\dps 
\Big\{ (p_2, s_i) \in {\mathbb C}^* \times {\mathbb C}^N \, \Big| \,
r - \sum_{i=1}^N |s_i|^2 + (N-k) |p_2|^2 = G_k (s_i) = 0
\Big\} 
\; .
$}
\end{split} \label{mfds}
\end{align}
Notice the following comments:
The LG superpotential $W_{\rm LG} = \vev{p_1} G_k (S)$ has an isolated
singularity at $S_i =0$, but $W_{\rm LG} = P_1 \cdot G_k (S)$ has no
isolated singularities.
Thus there may be no descriptions for (II-1) {\it as a minimal model}.
(There still exists a possibility that (II-1) could be described
as another, but unknown yet, well-defined CFT.)
The massless theories (I-2) and (II-2) 
are obtained by deformations of VEVs in (I-1) and (II-1), respectively.
When $k=2$, the massless theories (I-1) and (I-2) reduce
to the orbifolded sigma models on ${\cal M}^{p_1 \neq 0}$,
 because 
$W_{\rm LG} = \vev{p_1} G_{k=2}(S)$ is quadratic and generates mass terms of $S_i$.

In the classical level, 
there are no selection rules whether the four (or intrinsically two) theories 
will be realized as the true massless effective theory. 
However we can consider the T-dualized theories of them even in the
classical level. 


\vspace{3mm}

Now let us briefly discuss the T-dualized theory of the linear sigma
model.\footnote{The generic formulation of the T-dualized action are
  discussed in Ref. \cite{HV00}.}
Under T-duality, a chiral superfield $\Phi_a$ in the linear sigma model 
is replaced by a twisted chiral superfields $Y_a$ such as
$2 \ol{\Phi}_a \, \e^{2 Q_a V} \Phi_a = Y_a + \ol{Y}_{a}$.
Note that the imaginary part of $Y_a$ is periodic with period $2 \pi$. 
The T-dualized theory is described by an exact twisted superpotential
$\wt{W}$ and the period integral $\wh{\Pi}$ in terms of $Y_a$ and $\Sigma$:
\bsubeq
\begin{gather}
\text{\s $\dps
\wt{W} \ = \ \Sigma \Big( \sum_{i=1}^N Y_i - k Y_{P_1} - (N-k)
Y_{P_2} - t \Big) + \sum_{i=1}^N \e^{- Y_i} + \e^{- Y_{P_1}} + \e^{-
  Y_{P_2}} 
\; ,
$} \label{general-Wac} \\
\text{\s $\dps
\wh{\Pi} \ = \ \int \d \Sigma \prod_{i=1}^N \d Y_i \, \d Y_{P_1} \,
\d Y_{P_2} \, (k \Sigma) \, \exp \big( - \wt{W} \big)
\; . 
$} \label{period}
\end{gather}
\esubeq
In the IR limit $e \to \infty$, the dynamics of $\Sigma$ is frozen 
and thus the twisted superpotential reduces to $\wt{W} = \sum_{a} \e^{- Y_a}$ with
a constraint $\sum_a Q_a Y_a - t = 0$.
We always take this procedure when we analyze low energy effective
theories. 
Note that there exists the variable $k \Sigma$ in (\ref{period})
due to the existence of 
the superpotential $W_{\rm LSM}$ in the linear sigma model.

Here we construct mirror LG theories by
replacing $k \Sigma$ into $k \frac{\del}{\del t}$ and by performing 
field re-definitions.
First we consider the mirror LG theory of the effective theory (I-1) in
Figure \ref{vacuum-mfd}.
Re-defining $X_i \equiv \e^{- \frac{1}{k} Y_i}$ and 
$X_{P_2} \equiv \e^{\frac{N-k}{k} Y_{P_2}}$, 
we find that the measure in (\ref{period}) remains canonical and
the twisted LG superpotential becomes
\begin{align}
\text{\s $\dps
\Big\{ 
\wt{W}_k \ = \ X_1^k + \cdots + X_N^k 
+ X_{P_2}^{- \frac{k}{N-k}} + \e^{\frac{t}{k}} X_1 \cdots X_N X_{P_2}
\Big\} \Big/ ({\mathbb Z}_k)^{N}
\; .
$}
\label{LG-twist-On-CPnk-ksol}
\end{align}
Note that the orbifold symmetry $({\mathbb Z}_k)^{N}$,
derived from $Y_a \to Y_a + 2 \pi i$,  
acts on $X_i$ and $X_{P_2}$ such as
$X_i \to \omega_i X_i$ and $X_{P_2} \to \omega_{P_2} X_{P_2}$, 
where $\omega_i$ and $\omega_{P_2}$ satisfy 
$\omega_i^{k} = \omega_{P_2}^{- \frac{k}{N-k}} 
= \omega_1 \cdots \omega_N \omega_{P_2} = 1$. 
The negative power term $X_{P_2}^{- \frac{k}{N-k}}$ emerges due to the
existence of the noncompact direction in the original CY. 
The similar situation appears when we consider the LG theory of the
deformed conifold \cite{GV95} and so on.

We can also perform another field re-definition
$X_i \equiv \e^{- \frac{1}{N-k} Y_i}$ and $X_{P_1} \equiv
\e^{\frac{k}{N-k} Y_{P_1}}$ preserving a canonical measure in
(\ref{period}).
Due to this, we obtain another twisted LG superpotential 
\begin{align}
\text{\s $\dps
\Big\{ 
\wt{W}_{N-k} \ = \ X_1^{N-k} + \cdots + X_N^{N-k} 
+ X_{P_1}^{- \frac{N-k}{k}} + \e^{\frac{t}{N-k}} X_1 \cdots X_N X_{P_1}
\Big\} \Big/ ({\mathbb Z}_{N-k})^{N}
\; .
$}
\label{LG-twist-On-CPnk-nksol}
\end{align}
Note that the orbifold symmetry $({\mathbb Z}_{N-k})^{N}$
acts on $X_i$ and $X_{P_1}$ such as
$X_i \to \omega_i X_i$ and $X_{P_1} \to \omega_{P_1} X_{P_1}$, 
where $\omega_i$ and $\omega_{P_1}$ satisfy 
$\omega_i^{N-k} = \omega_{P_1}^{- \frac{N-k}{k}} 
= \omega_1 \cdots \omega_N \omega_{P_1} = 1$. 
By virtue of this orbifold symmetry, 
we can conjecture that this twisted theory is a mirror
LG of the low energy theory (II-1).

Replacing $k \Sigma$ into $\frac{\del}{\del Y_{P_1}}$ in (\ref{period}), we
obtain two CY geometries.
The first one is 
\begin{align}
\begin{split}
&\LS \ \ \text{\s $\dps 
\wt{\cal M}_{k} \ = \ 
\Big\{
(Z_i; u,v) \in {\mathbb C}^{N+2} \, \Big| \,
{\cal F} (Z_i) = 0 \; , \ {\cal G} (Z_b; u,v) = 0
\Big\} \Big/ \big\{ {\mathbb C}^* \times ({\mathbb Z_{k}})^{N-2} \big\}
\; , 
$} \\
& \ \ 
\text{\s $\dps
{\cal F} (Z_i) \ = \ \sum_{a=1}^k Z_a^{k} + \psi Z_1 \cdots Z_k 
\; , \ \ \
{\cal G} (Z_b; u, v) \ = \ \sum_{b=k+1}^N Z_b^{k} + 1 - uv 
\; , \ \ \ 
\psi \ = \ \e^{t/k} Z_{k+1} \cdots Z_N \; ,$}
\end{split} \label{mirror-O-CPNk-homo2}
\end{align}
where $u$ and $v$ are complex fields related to the noncompact
direction of the original CY ${\cal M}_{\rm CY}$;
$Z_i$ are related to $Y_i$ and $({\mathbb Z_k})^{N-2}$ acts on
$Z_i$; the ${\mathbb C}^*$ symmetry acts on the first $k$ fields $Z_a$. 
We can read this geometry as a CY geometry, 
i.e., a $({\mathbb Z}_k)^{N-2}$ orbifold CY hypersurface $\P{k-1}[k]$
parametrized by $\psi$ (this is represented by ${\cal F} = 0$)
under the constraint ${\cal G} =0$.
Similarly
we also obtain another CY geometry:
\begin{align}
\begin{split}
&\ls \ \ \text{\s $\dps
\wt{\cal M}_{N-k} \ = \ 
\Big\{
(Z_i; u,v) \in {\mathbb C}^{N+2} \, \Big| \,
F(Z_a) = 0 \; , \ G(Z_i; u, v) = 0
\Big\} \Big/ \big\{ {\mathbb C}^* \times ({\mathbb Z_{N-k}})^{N-2} \big\}
\; , 
$} \\
&\text{\s $\dps
F(Z_a) = \sum_{a=1}^k Z_a^{N-k} + 1 \; , \  
G(Z_i; u, v) = \sum_{b=k+1}^N Z_b^{N-k} + \psi Z_{k+1} \cdots Z_N 
\; , \ 
\psi = \big( \e^{\frac{t}{N-k}} - uv \big) Z_1 \cdots Z_k
\; .$}
\end{split} \label{mirror-O-CPNk-homo1}
\end{align}
This is a CY hypersurface $\P{N-k-1}[N-k]$
parametrized by $\psi$ 
(this is described by $G=0$) 
under the constraint $F =0$ with orbifold symmetry $({\mathbb Z}_{N-k})^{N-2}$.

\section{Discussions}

{}From the viewpoint of orbifold symmetries, 
it seems that the sigma models on (\ref{mirror-O-CPNk-homo2}) and
the sigma model on (\ref{mirror-O-CPNk-homo1}) 
are mirror duals of the theories (I-1) and (II-1),
respectively.
But it is now very hard to check this
conjecture and to determine the true mirror geometry of the original CY
because
we have not known how to define topological invariants on {\it noncompact} geometries. 
However, 
we might understand the true vacuum in the linear sigma model
and evaluate the correct LG theory and CY geometry in the T-dualized theory
when we evaluate free energies of the low energy effective theories
in Figure \ref{vacuum-mfd}.
We will explain the above discussions more precisely, 
and try to solve them in \cite{TK}.

\section{Acknowledgements}

The author would like to thank 
Hiroyuki Fuji,
Takahiro Masuda,
Shun'ya Mizoguchi,
Kazutoshi Ohta,
Hitoshi Sato 
and 
Dan Tomino
for valuable comments.
This work was supported in part by the JSPS Research Fellowships 
for Young Scientists (No.15-03926).

\bibliographystyle{plain}

}
\end{document}